\documentclass{article}

\usepackage{a4}

\scrollmode
\usepackage{amsmath}
\usepackage{amsfonts}
\usepackage{amssymb}
\usepackage{latexsym}
\usepackage{stmaryrd}
\usepackage{array}
\usepackage{exscale}
\newcommand{\nc}{\newcommand}
\newcommand{\ol}{\overline}
\newcommand{\ul}{\underline}
\newcommand{\es}{\emptyset}
\newcommand{\sm}{\setminus}
\newcommand{\ve}{\varepsilon}

\newcommand{\bc}{\bigcup}

\newcommand{\Lra}{\Leftrightarrow}

\newcommand{\ra}{\rightarrow}

\newcommand{\sse}{\subseteq}

\newcommand{\spe}{\supseteq}
\newcommand{\fa}{\forall}

\newcommand{\mr}{\mathrm}
\newcommand{\mc}{\mathcal}
\newcommand{\mf}{\mathfrak}

\newcommand{\DMO}{\DeclareMathOperator}
\newcommand{\DST}{\displaystyle}

\newcommand{\ZZ}{\mathbb{Z}}
\newcommand{\NN}{\mathbb{N}}
\newcommand{\NNZ}{\NN_0}

\newcommand{\RR}{\mathbb{R}}

\mathchardef\breakingcomma\mathcode`\,
{\catcode`,=\active
  \gdef,{\breakingcomma\discretionary{}{}{}}
}
\newcommand{\mathlist}[1]{$\mathcode`\,=\string"8000 #1$}
\usepackage{listings}
\newcommand{\inl}[1]{\lstinline$#1$}
\newcommand{\mb}{{\:|\:}} 
\newcommand{\set}[1]{\{ #1 \}}

\nc{\simlvi}[1]{\!\sim_{#1}}
\DeclareMathOperator{\addcup}{{\stackrel{\text{\raisebox{-2.2ex}[-0ex][-0ex]{\large$\cdot$}}}{\cup}}} 
\nc{\apprel}[3]{{#1}(#2)_{(#3)}} 
\nc{\cmpli}[1]{\complement^1_{#1}} 
\nc{\cmplzi}[1]{\complement^0_{#1}} 
\nc{\cmplzoi}[1]{\complement^*_{#1}} 

\nc{\zf}{\mr{ZF}}
\nc{\zfmf}{\zf^0} 
\nc{\zfc}{\mr{ZFC}}
\nc{\zfcmf}{\zfc^0} 
\nc{\bst}{\mr{BST}} 
\newcommand{\tb}[2]{\set{#1, \dots, #2}} 
\providecommand{\abs}[1]{\lvert #1 \rvert} 
\providecommand{\norm}[1]{\lVert #1 \rVert} 
\makeatletter
\DeclareRobustCommand{\genericinterval}[2]{%
  \@ifstar{\genericinterval@star{#1}{#2}}{\genericinterval@nostar{#1}{#2}}}
\newcommand{\genericinterval@star}[4]{\mathopen{}\mathclose{\left#1#3,#4\right#2}}
\newcommand{\genericinterval@nostar}[4]{\mathopen{#1}#3,#4\mathclose{#2}}

\makeatother
\nc{\untit}[2]{{#1}^{#2 \downarrow}} 
\nc{\obit}[2]{{#1}^{#2 \uparrow}} 
\nc{\inzEKi}[1]{\mc{I}^{\mr{V}}_{#1}}

\nc{\inzKEi}[1]{\mc{I}^{\mr{E}}_{#1}}

\nc{\adjEi}[1]{\mc{A}^{\mr{V}}_{#1}}

\nc{\BD}[1]{{#1}\text{-}\mr{BD}}

\nc{\konv}[2]{{#1}[{#2}]} 
\nc{\actpres}[1]{\Phi_{#1}} 
\newcommand{\floor}[1]{\lfloor #1 \rfloor}

\nc{\Prim}{\mc{PR}} 
\DeclareMathOperator{\ord}{ord}

\nc{\sselr}{\sse^{\mapsto}}
\nc{\sserl}{\sse^{\mapsfrom}}
\nc{\spelr}{\spe^{\mapsto}}
\nc{\sperl}{\spe^{\mapsfrom}}
\nc{\ball}[1]{\mr{B}^{#1}} 
\nc{\oball}[1]{\breve{\mr{B}}^{#1}} 
\nc{\pball}[1]{\dot{\mr{B}}^{#1}} 
\nc{\prr}[1]{\dot{\RR}^{#1}} 
\nc{\sph}[1]{\mr{S}^{#1}} 
\nc{\ssim}[1]{s\sigma_{#1}} 
\nc{\koerper}[1]{\norm{#1}}
\nc{\Ccovdim}{\mc{CD}}
\nc{\Cinddim}{\mc{SID}}

\nc{\CInddim}{\mc{LID}}

\DeclareMathOperator{\diffop}{D} 
\DeclareMathOperator*{\diffoplimit}{D} 
\nc{\diffopc}[1]{\sideset{_{#1}}{}\diffoplimit} 
\nc{\diffopp}[1]{\diffop_{#1}} 
\nc{\diffopcp}[2]{\sideset{_{#2}}{_{#1}}\diffoplimit} 
\nc{\meanH}[2]{\mf{M}_{#1,#2}} 
\nc{\emean}[2]{\mf{M}_{\exp_{#1},#2}} 
\DeclareMathOperator{\mor}{Mor}
\DeclareMathOperator{\Hom}{Hom} 
\nc{\autoerw}[1]{\mr{Aut}^{#1}} 
\nc{\komma}[2]{(#1 \downarrow #2)} 
\nc{\Kmat}{\mf{MAT}} 
\nc{\Khmat}{\mf{HMAT}} 
\nc{\homfun}[1]{\mor_{#1}(-_1,-_2)} 
\nc{\homfunae}[1]{\mor_{#1}(-_1)} 
\nc{\homfunbe}[1]{\mor_{#1}(-_2)} 
\nc{\homfunxy}[3]{\mor_{#1}(#2(-_1), #3(-_2))}
\nc{\homfunx}[2]{\mor_{#1}(#2(-_1), -_2)}
\nc{\homfuny}[2]{\mor_{#1}(-_1, #2(-_2))}
\nc{\homfuna}[2]{\mor_{#1}(#2, -)} 
\nc{\homfunb}[2]{\mor_{#1}(-, #2)} 
\nc{\hhomfuna}[2]{\Hom_{#1}(#2, -)} 
\nc{\hhomfunb}[2]{\Hom_{#1}(-, #2)} 
\newcommand{\Va}{\mc{V\hspace{-0.1em}A}}

\newcommand{\Lit}{\mc{LIT}}

\newcommand{\Cls}{\mc{CLS}}

\newcommand{\Sat}{\mc{SAT}}

\newcommand{\Usat}{\mc{USAT}}

\newcommand{\Musat}{\mc{M\hspace{0.8pt}U}} 
\newcommand{\Musati}[1]{\Musat_{\!#1}} 
\newcommand{\Smusat}{\mc{S}\Musat} 
\newcommand{\Smusati}[1]{\Smusat_{\!#1}}

\nc{\Clsoo}{\Cls^{1,1}} 

\DeclareMathOperator{\var}{var}

\newcommand{\Clash}{\mc{HIT}} 

\newcommand{\Uclash}{\mc{U}\Clash} 
\newcommand{\Uclashi}[1]{\Uclash_{\!\!#1}}

\newcommand{\Lean}{\mc{LEAN}}
\newcommand{\Leani}[1]{\Lean_{\!#1}}

\DeclareMathOperator{\dpl}{DP} 
\newcommand{\dpi}[1]{\dpl_{\!#1}}
\DMO{\premr}{ax} 
\DMO{\concr}{C} 
\DMO{\allcr}{cl} 

\DMO{\thardness}{thd} 
\DMO{\phardness}{phd} 
\DMO{\whardness}{awid} 
\DMO{\dep}{dep} 
\DMO{\hts}{hs} 
\DMO{\semspace}{css} 
\DMO{\resspace}{crs} 
\DMO{\treespace}{cts} 
\nc{\bth}[1]{\langle{#1}\rangle} 
\DMO{\rsub}{r_S} 
\DMO{\rk}{r} 
\DMO{\ro}{\rk_1} 
\DMO{\rki}{\rk_{\infty}} 
\DMO{\rpl}{r^{pl}} 
\DMO{\ropl}{\rk_1^{pl}} 
\nc{\rslur}{\xrightarrow{\text{SLUR}}} 
\nc{\rslurs}{\rslur_{\!*}} 
\DMO{\slur}{slur} 
\nc{\Slur}{\mc{SLUR}} 
\nc{\rkslur}[1]{\xrightarrow{\text{SLUR}_{#1}}} 
\nc{\rkslurs}[1]{\rkslur{#1}_{\!*}} 
\nc{\Altsluri}[1]{\Slur(#1)}
\nc{\Altslurstari}[1]{\Slur\text{\textasteriskcentered}(#1)}
\nc{\Canoni}[1]{\mr{CANON}(#1)}
\nc{\rkslurstar}[1]{\xrightarrow{\text{SLUR\textasteriskcentered}#1}} 
\nc{\rkslursstar}[1]{\rkslurstar{#1}_{\!*}} 
\DMO{\slurstar}{\slur\!\text{\textasteriskcentered}}
\nc{\Urefc}{\mc{UC}}
\nc{\Propc}{\mc{PC}}
\nc{\Wrefc}{\mc{WC}} 
\DeclareMathOperator{\vdeg}{vd} 
\DeclareMathOperator{\minvdeg}{\mu\!\vdeg} 
\DMO{\varmvd}{\var_{\minvdeg}} 
\DMO{\nfc}{fc} 
\DMO{\maxnfc}{\nu\!\nfc} 
\DeclareMathOperator{\nonmer}{nM} 
\nc{\svbf}{\mc{VB}} 
\nc{\svbfs}{\mc{VB}^*} 
\DMO{\potp}{pp} 
\DMO{\potprec}{NM} 
\DMO{\minnonmer}{VDM} 
\DMO{\minnonmerh}{VDH} 
\DMO{\maxsmar}{FCM} 
\DMO{\maxsmarh}{FCH} 
\DMO{\varsing}{\var_s} 
\DMO{\varosing}{\var_{1s}} 
\DMO{\varnosing}{\var_{\neg1s}} 
\nc{\Musatns}{\Musat'} 
\nc{\Musatnsi}[1]{\Musati{#1}'}
\nc{\Smusatns}{\Smusat'} 
\nc{\Smusatnsi}[1]{\Smusati{#1}'}
\nc{\Uclashns}{\Uclash'} 
\nc{\Uclashnsi}[1]{\Uclashi{#1}'}
\nc{\tsdp}{\xrightarrow{\text{sDP}}}
\nc{\tsdps}{\tsdp_{\!*}}
\nc{\tosdp}{\xrightarrow{\text{1sDP}}}
\nc{\tosdps}{\tosdp_{\!*}}
\DMO{\sdp}{sDP} 
\DMO{\osdp}{sDP_1} 
\nc{\cflmusat}{\mc{CF}\Musat} 
\nc{\cflmusati}[1]{\mc{CF}\Musati{#1}}
\nc{\cflimusat}{\mc{CFI}\Musat} 
\DMO{\sNF}{sNF} 
\DMO{\eqp}{eqp} 
\DMO{\sgp}{sp} 
\DMO{\singind}{si} 
\DMO{\osingind}{si_1} 
\DMO{\shyp}{svh} 
\DMO{\sdph}{ssh} 
\DMO{\msdph}{mss} 
\DMO{\osdph}{ssh_1} 
\DMO{\mosdph}{mss_1} 
\DMO{\mps}{mps} 
\DMO{\purec}{puc} 
\DMO{\doping}{D}
\nc{\glue}[4]{\operatorname{glue}((#1,#2), (#3,#4))} 
\nc{\gluea}[3]{#1 \mathbin{\boxplus}_{#3} #2} 
\DMO{\frl}{fl} 
\nc{\Con}{\mr{Con}}
\nc{\Log}{\mr{Log}}
\nc{\Lin}{\mr{Lin}}
\nc{\Pol}{\mr{Pol}}
\nc{\ExL}{\mr{ExL}}
\nc{\ExP}{\mr{ExP}}
\nc{\CTime}{\mr{CTime}}
\nc{\CSpace}{\mr{CSpace}}
\nc{\LTime}{\mr{LTime}}
\nc{\LSpace}{\mr{L}}
\nc{\NLSpace}{\mr{NL}}
\nc{\LinTime}{\mr{LinTime}}
\nc{\LinSpace}{\mr{LinSpace}}
\nc{\PTime}{\mr{P}}
\nc{\PSpace}{\mr{PSpace}}
\nc{\Np}{\mr{NP}}
\nc{\Conp}{\text{coNP}}
\nc{\NPSpace}{\mr{NPSpace}}
\nc{\CoNPSpace}{\mr{coNPSpace}}
\nc{\ELTime}{\mr{ELTime}}
\nc{\ELSpace}{\mr{ELSpace}}
\nc{\EPTime}{\mr{EPTime}}
\nc{\EPSpace}{\mr{EPSpace}}
\nc{\NEPTime}{\mr{NEPTime}}
\nc{\polydelta}[1]{\Delta_{#1}^{\mr P}}
\nc{\polypi}[1]{\Pi_{#1}^{\mr P}}
\nc{\polysigma}[1]{\Sigma_{#1}^{\mr P}}
\nc{\Ph}{\mr{PH}}

\nc{\Dp}{D^P}
\nc{\PllC}[2]{{\text{$\mr{PT}$/$\mr{WK}$}(#1, #2)}} 
\nc{\Nc}{\mr{NC}}
\nc{\Nci}[1]{\Nc^{#1}}
\nc{\Ac}{\mr{AC}}
\nc{\Aci}[1]{\Ac^{#1}}
\nc{\pmodpoly}{P / \mathrm{poly}}
\nc{\Wh}[1]{\mr{W}[#1]} 
\nc{\Rl}{\mr{RL}}
\nc{\coRl}{\mr{coRL}}
\nc{\Rp}{\mr{RP}}
\nc{\coRp}{\mr{coRP}}
\nc{\Zpp}{\mr{ZPP}}
\nc{\Bpp}{\mr{BPP}}
\nc{\Pp}{\mr{PP}}
\nc{\Reach}{\mr{STCON}} 
\nc{\Undreach}{\mr{USTCON}} 
\nc{\Pcol}[2]{\mr{COL}(#1,#2)} 
\nc{\Pscol}[2]{\mr{SCOL}(#1,#2)} 
\nc{\Psorcol}[2]{\mr{SORCOL}(#1,#2)} 
\DMO{\slp}{slp}
\nc{\Mss}{\mr{MSS}}
\nc{\Key}{\mr{KEY}}
\nc{\Keyi}[1]{\Key_{\!#1}}
\nc{\Nbmss}{N_{\mr{bm}}} 
\nc{\Nbkey}{N_{\mr{bk}}} 
\nc{\Rnb}{N_{\mr{b}}}
\nc{\Rnk}{N_{\mr{k}}}
\nc{\Rnr}{N_{\mr{r}}}

\nc{\Byte}{\mr{B}[8]}
\nc{\QByte}{\mr{B}[4,8]}
\nc{\KByte}{\mc{B}} 
\nc{\RQByte}{\mc{QB}} 

\nc{\ramz}[3]{\mr{ram}_{#1}^{#2}(#3)} 
\nc{\waez}[2]{\mr{vdw}_{#1}(#2)} 
\nc{\gtz}[2]{\mr{grt}_{#1}(#2)} 
\nc{\pdwaez}[2]{\mr{vdw}_{#1}^{\mr{pd}}(#2)} 
\nc{\absfeh}[1]{\delta_{#1}} 
\nc{\relfeh}[1]{\ve_{#1}} 
\usepackage{theorem} 
\usepackage{hyperref} 
\newtheorem{defi}{Definition}[section]
\newtheorem{lem}[defi]{Lemma}
\newtheorem{thm}[defi]{Theorem}
\newtheorem{corol}[defi]{Corollary}

\newtheorem{conj}[defi]{Conjecture}

\newtheorem{examp}[defi]{Example}

\theorembodyfont{\rmfamily}

\theorembodyfont{}
\newenvironment{prf}{\noindent\textbf{Proof:}\;}{\par\noindent\ignorespacesafterend}

\newcommand{\Qed}{\hfill $\square$}

\newcounter{dDef} 

\newcounter{dLem} 

\newcounter{dThm} 

\newcounter{dPro} 

\newcounter{Beispielzaehler}

\nc{\bm}{\boldmath}
\nc{\bmm}[1]{\mbox{\bm$\DST #1$}}
\nc{\mi}[1]{\bmm{\mathrm{(#1):}} \quad}
\usepackage{enumerate}
\usepackage[active]{srcltx}

\newcommand{\Schrift}{report}
\newcommand{\Liste}{SAT \sep minimal unsatisfiability \sep hitting clause-sets \sep orthogonal DNF \sep disjoint DNF \sep variable degree \sep minimum variable degree in CNF \sep number of full clauses in CNF \sep deficiency \sep full subsumption resolution \sep Smarandache primitive numbers \sep meta-Fibonacci sequences \sep non-Mersenne numbers}

\providecommand{\keywords}[1]{\textbf{\textit{Keywords}} #1}
\providecommand{\sep}{, }

\DeclareMathOperator{\ruler}{ru}
\newcommand{\ismar}{\operatorname{i}_{\mathrm{S}}} 
\newcommand{\ssmar}{\operatorname{sl}_{\mathrm{S}}} 

\DMO{\platop}{\mf{P}}

\newcommand{\Esnm}{\mc{SN\!M}}

\begin{document}

\title{Parameters for minimal unsatisfiability:\\Smarandache primitive numbers and full clauses}

\author{
  \href{http://cs.swan.ac.uk/~csoliver}{Oliver Kullmann}\\
  Computer Science Department\\
  Swansea University\\
  Swansea, SA2 8PP, UK
  \and
  \href{http://logic.sysu.edu.cn/faculty/zhaoxishun/en/}{Xishun Zhao}\thanks{This research was partially supported by NSFC Grand 61272059, NSSFC Grant 13\&ZD186 and MOE Grant 11JJD7200020.}\\
  Institute of Logic and Cognition\\
  Sun Yat-sen University\\
  Guangzhou, 510275, P.R.C.
}

\maketitle

\begin{abstract}
  We establish a new bridge between propositional logic and elementary number theory. A \emph{full clause} in a conjunctive normal form (CNF) contains all variables, and we study them in minimally unsatisfiable clause-sets (MU); such clauses are strong structural anchors, when combined with other restrictions. Counting the maximal number of full clauses for a given \emph{deficiency} $k$, we obtain a close connection to the so-called ``Smarandache primitive number'' $S_2(k)$, the smallest $n$ such that $2^k$ divides $n!$.

The deficiency $k \ge 1$ of an MU is the difference between the number of clauses and the number of variables. We also consider the subclass UHIT of MU given by unsatisfiable hitting clause-sets, where every two clauses clash. While MU corresponds to irredundant (minimal) covers of the boolean hypercube $\set{0,1}^n$ by sub-cubes, for UHIT the covers must indeed be partitions.

We study the four fundamental quantities $\maxsmarh, \maxsmar, \minnonmerh, \minnonmer: \NN \ra \NN$, defined as the maximum number of \ul{f}ull \ul{c}lauses in UHIT resp.\ MU, resp.\ the maximal minimal number of occurrences of a variable (the \ul{v}ariable \ul{d}egree) in UHIT resp.\ MU, in dependency on the deficiency. We have the relations $\maxsmarh(k) \le \maxsmar(k) \le \minnonmer(k)$ and $\maxsmarh(k) \le \minnonmerh(k) \le \minnonmer(k)$, together with $\minnonmer(k) \le \nonmer(k) \le k + 1 + \log_2(k)$, for the ``non-Mersenne numbers'' $\nonmer(k)$, enumerating the natural numbers except numbers of the form $2^n - 1$.

We show the lower bound  $S_2(k) \le \maxsmarh(k)$; indeed we conjecture this to be exact. The proof rests on two methods: Applying an expansion process, fundamental since the days of Boole, and analysing certain recursions, combining an application-specific recursion with a recursion from the field of \emph{meta-Fibonacci sequences}.

The $S_2$-lower bound together with the $\nonmer$-upper-bound yields a good handle on the four fundamental quantities, especially for those $k$ with $S_2(k) = \nonmer(k)$ (we show there are infinitely many such $k$), since then the four quantities must all be equal to $S_2(k) = \nonmer(k)$. With the help of this we determine them for $1 \le k \le 13$.
\end{abstract}

\keywords{\Liste}

\tableofcontents

\section{Introduction}
\label{sec:intro}

We study combinatorial parameters of conjunctive normal forms (CNFs) $F$, conjunctions of disjunctions of literals, under the viewpoint of extremal combinatorics: We maximise the number of ``full clauses'' in $F$ for a given ``deficiency'' $\delta(F)$, where not all $F$ are considered (that number would not be bounded), but only ``minimally unsatisfiable'' $F$. We use exact methods, establishing links to elementary number theory and to the theory of special recursions.

To help the reader, we give now the definitions, in a somewhat unusual way, which is nevertheless fully precise. CNFs as combinatorial objects are ``clause-sets'', where for this introduction we just use natural numbers (positive integers) as logical ``variables''. More precisely, we consider non-zero integers as literals $x$ with arithmetical negation $-x$ the logical negation, while clauses are finite sets $C$ of Literals (non-zero integers), such that for $x \in C$ we don't have $-x \in C$ (logically speaking, $C$ must not be tautological), and clause-sets $F$ are finite sets of clauses. The set $\var(F)$ of variables of $F$ is the set of absolute values of literals occurring in $F$. A \emph{full clauses} $C \in F$ is a clause of maximal possible length, that is, of length $\abs{\var(F)}$, in other words, all variables must occur in $C$ (negated or unnegated); the number of full clauses of $F$ is denoted by $\nfc(F)$. A clause-set $F$ is satisfiable iff there exists a clause $C$ (which represents the set of ``literals set to true''), which intersects all clauses of $F$ (note that this is non-trivial, since $C$ must not contain complementary literals $x$ and $-x$), otherwise $F$ is unsatisfiable. Moreover, unsatisfiable $F$, where removal of any clause makes them satisfiable, are called \emph{minimally unsatisfiable}, while the set of all of them is denoted by $\Musat$. The main parameter is the deficiency $\delta(F) = c(F) - n(F) \in \ZZ$, where $c(F) := \abs{F}$ is the number of clauses, and $n(F) := \abs{\var(F)}$ is the number of variables. The most basic result of the field, ``Tarsi's Lemma'' (\cite{AhLi86}), states $\delta(F) \ge 1$ for $F \in \Musat$. An example of an unsatisfiable clause-set is $\set{\set{-1},\set{1},\set{1,2}}$, which is not minimal, but $F_1 := \set{\set{-1},\set{1}} \in \Musat$, with $\delta(F_1) = 2 - 1 = 1$ and $\nfc(F_1) = 2$. An example of $F \in \Musat$ with $\nfc(F) = 0$ is $F_2 := \set{\set{-1,2},\set{-2,3},\set{-3,1},\set{1,2},\set{-2,-3}}$, where $\delta(F_2) = 2$. Indeed we mainly concentrate on a subset of $\Musat$, namely $\Uclash \subset \Musat$, the \emph{unsatisfiable hitting clause-sets}, given by those $F \in \Musat$ such that for each $C, D \in F$, $C \ne D$, there is a ``clash'', that is, there is $x \in C$ with $-x \in D$. We have $F_1 \in \Uclash$ and $F_2 \notin \Uclash$; the latter can be ``repaired'' with $F_3 := \set{\set{-1,2},\set{-2,3},\set{-3,1},\set{1,2,3},\set{-1,-2,-3}} \in \Uclash$ (still $\delta(F_3) = 2$, but now $\nfc(F_3) = 2$).

Now we denote by $\maxsmar(k)$ the maximum of $\nfc(F)$ for $F \in \Musat$ with $\delta(F) = k$ (short: $F \in \Musati{\delta=k}$). From \cite[Theorem 15]{KullmannZhao2011Bounds} follows the upper bound $\maxsmar(k) \le \nonmer(k)$ for the \emph{non-Mersenne numbers} $\nonmer(k) \in \NN$, with $k + \floor{\log_2(k+1)} \le \nonmer(k) \le k + 1 + \floor{\log_2(k)}$ (\cite[Corollary 10]{KullmannZhao2011Bounds}). Until now no general lower bound on $\maxsmar(k)$ was known, and we establish $S_2(k) \le \maxsmar(k)$. Here $S_2(k)$, as introduced in \cite{Smarandache1993OPNS}, is the smallest $n \in \NNZ$ such that $2^k$ divides $n!$, and various number-theoretical results on $S_2$ and the generalisation $S_p$ for prime numbers $p$ are known. Actually we show a stronger lower bound, namely we do not consider all $F \in \Musati{\delta=k}$, but only those $F \in \Uclash$, yielding $\maxsmarh(k)$ with $\maxsmarh(k) \le \maxsmar(k)$, and we show $S_2 \le \maxsmarh$. The elements of $\Uclash$  are known in the DNF language as ``orthogonal'' or ``disjoint'' tautological DNF, and when considering arbitrary boolean functions, then also ``disjoint sums of products'' (DSOP) or ``disjoint cube representations'' are used;  see \cite[Section 4.4]{Schneeweiss1989BoolescheFunktionen} or \cite[Chapter 7]{CramaHammer2011BooleanFunctions}.

\subsection{Background}
\label{sec:introback}

The central underlying research question is the programme of classification of $\Musat$ in the deficiency, that is, the characterisation of the layers $\Musati{\delta=k}$ for $k \in \NN$. A special case of the general classification is the classification of $\Uclashi{\delta=k}$. The earliest source \cite{AhLi86} showed (in modern notation) $\delta(F) \ge 1$ for $F \in \Musat$, and characterised the special case $\Smusati{\delta=1} \subset \Musati{\delta=1}$, where $\Smusat \subset \Musat$ contains those $F \in \Musat$ such that no literals can be added to any clauses without destroying unsatisfiability. Later \cite{DDK98} characterised $\Musati{\delta=1}$ via matrices, while the intuitive characterisation via binary trees was given in \cite[Appendix B]{Ku99dKo}, where also $\Smusati{\delta=1} = \Uclashi{\delta=1}$ has been noted. In the form of ``$S$-matrices'', the class $\Musati{\delta=1}$ had been characterised earlier in \cite{KLM1984,Klee1987RecursiveStructure}, going back to a conjecture on Qualitative Economics (\cite{Gorman1964QualitativeEconomics}), and where the connections to this field of matrix analysis, called ``Qualitative Matrix Analysis (QMA)'', were first revealed in \cite{Ku00f} (see \cite[Subsection 11.12.1]{Kullmann2007HandbuchMU} and \cite[Subsection 1.6.4]{KullmannZhao2010Extremal} for overviews). Another proof of $\delta(F) \ge 1$ for $F \in \Musat$ is obtained as a special case of \cite[Corollary 4]{BergerFelzenbaumFraenkel1988Covers}, as pointed out in \cite{BergerFelzenbaumFraenkel1990Covers}.

$\Smusati{\delta=2}$ and partially $\Musati{\delta=2}$ were characterised in \cite{KleineBuening2000SubclassesMU}, with further information on $\Musati{\delta=2}$ in \cite{KullmannZhao2012ConfluenceJ}. \cite{FKS00} showed that all layers $\Musati{\delta=k}$ are poly-time decidable.

A key element for these investigations into the structure of $\Musat$ is the \emph{min-var-degree} $\minvdeg(F) := \min_{v \in \var(F)} \abs{\set{C \in F : \set{-v,v} \cap C \ne \es}}$, the minimal variable-degree of $F$, and its maximum $\minnonmer(k)$ over all $F \in \Musati{\delta=k}$. Indeed the key to the characterisation of $\Musati{\delta=1}$ in \cite{DDK98} as well as in \cite{KLM1984} was the proof of $\minnonmer(1) = 2$. The first general upper bound $\fa\, k \in \NN : \minnonmer(k) \le 2 k$ was shown in \cite[Lemma C.2]{Ku99dKo}. Now in \cite{KullmannZhao2011Bounds}, mentioned above, we actually showed the upper bound $\minnonmer(k) \le \nonmer(k)$. Using $\nfc(F)$ for the number of full clauses in $F$, obviously $\nfc(F) \le \minvdeg(F)$ holds. $\maxsmar(k)$ is the maximum of $\nfc(F)$ over all $F \in \Musati{\delta=k}$, thus $\maxsmar(k) \le \minnonmer(k)$.

In \cite[Section 14]{KullmannZhao2010Extremal} we improve the upper bound to $\minnonmer \le \nonmer_1$, based on two results: $\minnonmer(6) = \nonmer(6) - 1 = 8$, and a recursion scheme, transporting this improvement to higher deficiencies, obtaining $\nonmer_1$ from $\nonmer$, where for infinitely many $k$ holds $\nonmer_1(k) = \nonmer(k) - 1$. The proof of $\minnonmer(6) = 8$ contains an application of full clauses, namely we use $\maxsmar(3) = 4$.

For the variation $\minnonmerh(k) \le \minnonmer(k)$, which only considers hitting clause-sets, we conjecture $\minnonmerh(k) = \minnonmer(k)$ for all $k \ge 1$. Furthermore we conjecture $\maxsmar(k) \ge \nonmer(k) - 1$, and thus the quantities $\nonmer(k), \minnonmer(k), \minnonmerh(k), \maxsmar(k)$ are believed to have at most a distance of $1$ to each other. On the other hand we conjecture $\maxsmarh(k) = S_2$, where $S_2(k)$ oscillates between the linear function $k+1$ and the quasi-linear function $k+1+\floor{\log_2(k)}$. Altogether the ``four fundamental quantities'' $\maxsmarh, \maxsmar, \minnonmerh, \minnonmer$ seem fascinating and important structural parameters, whose study continues to reveal new and surprising aspects of $\Musat$ and $\Uclash$; see Section \ref{sec:conclusion} for some final remarks.

It is also possible to go beyond $\Musat$: in \cite[Section 9]{KullmannZhao2010Extremal} it is shown that when considering the maximum of $\minvdeg(F)$ over all $F \in \Leani{\delta=k} \supset \Musati{\delta=k}$, the set of all ``lean'' clause-sets, that then $\nonmer(k)$ is the precise maximum for all $k \ge 1$. Lean clause-sets were introduced in \cite{Ku98e} as the clause-sets where it is not possible to satisfy some clauses while not touching the other clauses, and indeed were already introduced earlier, as ``non-weakly satisfiable formulas (matrices)'' in the field of QMA by \cite{KleeLadner1981WeakSAT}. Furthermore it is shown in \cite[Section 10]{KullmannZhao2010Extremal}, that there is a polytime ``autarky reduction'', removing some clauses which can be satisfied without touching the other clauses, which establishes for arbitrary clause-sets $F$ the upper bound $\minvdeg(F) \le \nonmer(\delta(F))$; an interesting open question here is to find the witnessing autarky in polynomial time.

\subsection{The lower bound}
\label{sec:introlower}

Back to the main result of this \Schrift, the proof of $S_2 \le \maxsmarh$ is non-trivial. Indeed the proof is relatively easy for a function $S'_2(k)$ defined by an appropriate recursion, motivated by employing ``full subsumption extension'' $C \leadsto C \cup \set{v}, C \cup \set{\ol{v}}$ in an optimal way. Then the main auxiliary result is $S'_2 = S_2$. For that we use another function, namely $a_2(k)$ as considered in \cite{RuskeyDeugau2009MetaFibonacci} in a more general form, while $a_2$ was introduced with a small modification in \cite{Conolly1989MetaFibonacci}. These considerations belong to the field of \emph{meta-Fibonacci sequences}, where special nested recursions are studied, initiated by \cite[Page 145]{Hofstadter1979GEB}. Via a combinatorial argument we derive such a nested recursion from the course-of-value recursion for $S'_2$, which yields $S'_2 = 2 a_2$. We also show $2 a_2 = S_2$ (this equality was conjectured on the OEIS \cite{Sloane2008OEIS}), and we obtain $S_2' = S_2$.

We obtain the inequality $S_2 \le \nonmer$, with the four fundamental quantities sandwiched inbetween. The deficiencies $k$ where equality holds are collected in the set $\Esnm$, which we show has infinitely many elements. For the elements $k$ of $\Esnm$, the four fundamental quantities coincide with $S_2(k) = \nonmer(k)$, which yields islands of precise knowledge about the four quantities. We apply this knowledge to determine the four quantities for $1 \le k \le 13$.

\subsection{Overview on results}
\label{sec:introover}

The main results of this \Schrift{} are as follows.
Theorem \ref{thm:Spad} proves $S_2 = 2 a_2$. Theorem \ref{thm:nestrecS'} shows a meta-Fibonacci recursion for $S'_2$, where $S'_2$ is introduced by a recursion directly related to our application. Theorem \ref{thm:smarecfac} then proves $S'_2 = S_2$. After these number-theoretic preparations, we consider subsumption resolution and its inversion (extension); Theorem \ref{thm:smalbfccon} combines subsumption extension and the recursion machinery, and shows $S_2 \le \maxsmarh$. In the remainder of the \Schrift{}, this fundamental result is applied. Theorem \ref{thm:S2upper} proves a tight upper bound on $S_2$, while Theorem \ref{thm:meanesnm} considers the cases where the lower bound via $S_2$ and the upper bound via $\nonmer$ coincides. Finally in Theorem \ref{thm:13values} we determine the four fundamental quantities for $1 \le k \le 13$ (see Table \ref{tab:valuesmaxnfc}).


\section{Preliminaries}
\label{sec:Preliminaries}

We use $\ZZ$ for the set of integers, $\NNZ := \set{n \in \ZZ : n \ge 0}$, and $\NN := \NNZ \sm \set{0}$. For maps $f, g: X \ra \ZZ$ we write $f \le g$ if $\fa\, x \in X : f(x) \le g(x)$.

On the set $\Lit$ of ``literals'' we have complementation $x \in \Lit \mapsto \ol{x} \in \Lit$, with $\ol{x} \ne x$ and $\ol{\ol{x}} = x$. We assume $\ZZ \sm \set{0} \sse \Lit$, with $\ol{z} = -z$ for $z \in \ZZ \sm \set{0}$. ``Variables'' $\Va \subset \Lit$ with $\NN \sse \Va$ are special literals, and the underlying variable of a literal is given by $\var: \Lit \ra \Va$, such that for $v \in \Va$ holds $\var(v) = \var(\ol{v}) = v$, while for $x \in \Lit \sm \Va$ holds $\ol{x} = \var(x)$. For a set $L \sse \Lit$ we define $\ol{L} := \set{\ol{x} : x \in L}$. A clause is a finite set $C$ of literals with $C \cap \ol{C} = \es$ ($C$ is clash-free). A clause-set is a finite set of clauses, the set of all clause-sets is $\Cls$.

For a clause $C$ we define $\var(C) := \set{\var(x) : x \in C} \subset \Va$, and for a clause-set $F$ we define $\var(F) := \bc_{C \in F} \var(C) \subset \Va$. We use the measure $n(F) := \abs{\var(F)} \in \NNZ$ and $c(F) := \abs{F} \in \NNZ$, while the deficiency is $\delta(F) := c(F) - n(F) \in \ZZ$.

The set of satisfiable clause-sets is denoted by $\Sat \subset \Cls$, which is the set of clause-sets $F$ such that there is a clause $C$ which intersects all clauses of $F$, i.e., with $\fa\, D \in F : C \cap D \ne \es$; the unsatisfiable clause-sets are $\Usat := \Cls \sm \Sat$.

The set $\Musat \subset \Usat$ of minimally unsatisfiable clause-sets is the set of $F \in \Usat$, such that for $F' \subset F$ holds $F' \in \Sat$. The unsatisfiable hitting clause-sets are given by $\Uclash := \set{F \in \Usat \mb \fa\, C, D \in F, C \ne D : C \cap \ol{D} \ne \es}$. It is easy to see that $\Uclash \subset \Musat$ holds, and that for all $F \in \Uclash$ holds $\sum_{C \in F} 2^{-\abs{C}} = 1$. While all definitions are given in this \Schrift{}, for some more background see \cite{Kullmann2007HandbuchMU}.

\subsection{Full clauses}
\label{sec:prelimfullclauses}

A \textbf{full clause} for $F \in \Cls$ is some $C \in F$ with $\var(C) = \var(F)$ (equivalently, $\abs{C} = n(F)$), and the number of full clauses is counted by $\nfc: \Cls \ra \NNZ$, which can be defined as $\bmm{\nfc(F)} := c(F \cap A(\var(F)))$, and where $\bmm{A(V)} \in \Uclash$ for some finite $V \subset \Va$ is the set of all clauses $C$ with $\var(C) = V$. Standardised versions of the $A(V)$ are $\bmm{A_n} := A(\tb 1n)$ for $n \in \NNZ$.

\begin{examp}\label{exp:An}
  In general $n(A_n) = n$, $c(A_n) = 2^n$ and $\delta(A_n) = 2^n-n$. Initial cases are $A_0 = \set{\bot}$, $A_1 = \set{\set{1},\set{-1}}$ and $A_2 = \set{\set{-1,-2},\set{-1,2},\set{1,-2},\set{1,2}}$.
\end{examp}

The following observation is contained in the proof of \cite[Utterly Trivial Observation]{Zeilberger2001CoveringSystems}:

\begin{lem}\label{lem:evenfulluhit}
  For $F \in \Uclash$, $F \not= \set{\bot}$, the number $\nfc(F)$ of full clauses is even.
\end{lem}
\begin{prf}
Let $n := n(F)$. We have $\sum_{C \in F} 2^{n-\abs{C}} = 2^n$, and thus $\sum_{C \in F} 2^{n-\abs{C}}$ is even (due to $n > 0$). Since $\sum_{C \in F, \abs{C} \not= n} 2^{n-\abs{C}}$ is even, the assertion follows. \Qed
\end{prf}

\subsection{The four fundamental quantities}
\label{sec:prelimfour}

For $F \in \Cls$ we define the var-degree as $\vdeg_F(v) := c(\set{C \in F : v \in \var(C)}) \in \NNZ$ for $v \in \Va$, while in case of $\var(F) \ne \es$ (i.e., $F \notin \set{\top, \set{\bot}}$) we define the min-var-degree $\bmm{\minvdeg(F)} := \min_{v \in \var(F)} \vdeg_F(v) \in \NN$.

\begin{defi}\label{def:fourquant}
  For $k \in \NN$ let
  \begin{itemize}
  \item $\bmm{\maxsmarh(k)} \in \NN$ be the maximal $\nfc(F)$ for $F \in \Uclashi{\delta=k}$;
  \item $\bmm{\maxsmar(k)} \in \NN$ be the maximal $\nfc(F)$ for $F \in \Musati{\delta=k}$;
  \item $\bmm{\minnonmerh(k)} \in \NN$ be the maximal $\minvdeg(F)$ for $F \in \Uclashi{\delta=k}$;
  \item $\bmm{\minnonmer(k)} \in \NN$ be the maximal $\minvdeg(F)$ for $F \in \Musati{\delta=k}$.
  \end{itemize}
  For $k=1$ the case $F = \set{\bot}$ is excluded in the last two definitions.
\end{defi}

By \cite[Lemma 9, Corollary 10, Theorem 15]{KullmannZhao2011Bounds}:
\begin{thm}[\cite{KullmannZhao2011Bounds}]\label{thm:uppernM}
  $\minnonmer(k) \le \nonmer(k) = k + \floor{\log_2(k+1+\floor{\log_2(k+1)})} \le k + 1 + \floor{\log_2(k)}$ for all $k \in \NN$.
\end{thm}
Here $\nonmer: \NN \ra \NN$ is the enumeration of natural numbers excluding the Mersenne numbers $2^n - 1$ for $n \in \NN$; the list of initial values is \mathlist{2,4,5,6,8,9,10,11,12,13,14,16,17} (\url{http://oeis.org/A062289}). In \cite[Theorem 14.4]{KullmannZhao2010Extremal} it is shown that $\minnonmer(6) = 8 = \nonmer(6) - 1$, extending this to an improved upper bound $\minnonmer \le \nonmer_1$ (\cite[Theorem 14.6]{KullmannZhao2010Extremal}, where $\nonmer_1: \NN \ra \NN$ can be defined as follows: $\nonmer_1(k) := \nonmer(k)$ for $k \in \NN$ with $k \ne 2^n - n + 1$ for some $n \ge 3$, while $\nonmer_1(2^n - n + 1) := \nonmer(2^n - n + 1) - 1 = 2^n$; see Table \ref{tab:valuesmaxnfc} for initial values.
\begin{thm}[\cite{KullmannZhao2010Extremal}]\label{thm:uppernM1}
  For $k \in \NN$ holds $\minnonmer(k) \le \nonmer_1(k) \le \nonmer(k)$.
\end{thm}

We conclude these preparations with a special property of $\maxsmarh(k)$ (supporting our Conjecture \ref{con:S2exact} that $\maxsmarh = S_2$), namely by Lemma \ref{lem:evenfulluhit} we have:
\begin{corol}\label{cor:maxsmarheven}
  $\maxsmarh(k)$ is even for all $k \in \NN$.
\end{corol}

\section{Some integer sequences}
\label{sec:intseq}

We review the ``Smarandache primitive numbers'' $S_2(k)$ and the meta-Fibonacci sequences $a_2(k)$. We show in Theorem \ref{thm:Spad}, that $S_2 = 2 a_2$ holds.

\subsection{Some preparations}
\label{sec:platseq}

We define two general operations $a \mapsto \Delta a$ and $a \mapsto \platop a$ for sequences $a$. First the (standard) $\Delta$-operator:
\begin{defi}\label{def:Delta}
  For $a: I \ra \ZZ$, where $I \sse \ZZ$ is closed under increment, we define $\bmm{\Delta a}: I \ra \ZZ$ by $\Delta a(k) := a(k+1) - a(k)$.
\end{defi}
So $a$ is monotonically increasing iff $\Delta a \ge 0$, while $a$ is strictly monotonically increasing iff $\Delta a \ge 1$. Sequences with exactly two different $\Delta$-values, where one of these values is $0$, play a special role for us, and we call them ``$d$-Delta'', where $d$ is the other value:
\begin{defi}\label{def:bidelta}
  A sequence $a: \NNZ \ra \ZZ$ is called \textbf{$d$-Delta} for $d \in \ZZ \sm \set{0}$, if $\Delta a(\NNZ) = \set{\Delta a(n)}_{n \in \NNZ} = \set{0,d}$.
\end{defi}

While the $\Delta$-operator determines the change to the next value, the \emph{plateau-operator} determines subsequences of unchanging values:
\begin{defi}\label{def:Fop}
  For a sequence $a: \NN \ra \ZZ$ which is non-stationary (for all $i$ there is $j > i$ with $a_j \ne a_i$) we define $\platop a : \NN \ra \NN$ (the ``plateau operator'') by letting $\platop a(n)$ for $n \in \NN$ be the size of the $n$-th (maximal) plateau of equal values (maximal intervals of $\NN$ where $a$ is constant).
\end{defi}
So $\platop a(1)$ is the size of the first plateau, $\platop a(2)$ the size of the second plateau, and so on; $\fa\, i \in \NN: a(i) \ne a(i+1)$ iff $\platop a$ is the constant $1$-function. For a $d$-Delta sequence $a$ from $\platop a$ and the initial value $a_1$ we can reconstruct $a$.

\subsection{Smarandache primitive numbers}
\label{sec:prelimsma}

The ``Smarandache Primitive Numbers'' were introduced in \cite[Unsolved Problem 47]{Smarandache1993OPNS}:
\begin{defi}\label{def:smafac}
  For $k \in \NNZ$ let $S_2(k)$ be the smallest $n \in \NNZ$ such that $2^k$ divides $n!$. Using $\ord_2(n)$, $n \in \NN$, for the maximal $m \in \NNZ$ such that $2^m$ divides $n$, we get that $S_2(k)$ for $k \in \NNZ$ is the smallest $n \in \NNZ$ such that $k \le \sum_{i=1}^n \ord_2(i)$.
\end{defi}
So $S_2(0) = 0$, and $\Delta S_2(\NNZ) = \set{0,2}$.

\begin{examp}\label{exp:Sp}
  $S_2(2) = S_2(3) = 4$, while $S_2(4) = 6$, since $\ord_2(1) = \ord_2(3) = 0$, while $\ord_2(2) = 1$ and $\ord_2(4) = 2$.
\end{examp}

The following is well-known and easy to show (see Subsection III.1 in \cite{Ibstedt1998ComputerNumber} for basic properties of $S_2(k)$):
\begin{lem}\label{lem:S_pseq}
  The sequence $S_2(1),S_2(2),S_2(3), \dots$ is obtained from the sequence $1,2,3,\dots$ of natural numbers, when each element $n \in \NN$ is repeated $\ord_2(n)$ many times.
\end{lem}

\begin{examp}\label{exp:Spseq}
  The numbers $S_2(k)$ for $k \in \tb{1}{25}$ are \mathlist{2,4,4,6,8,8,8,10,12,12,14,16,16,16,16,18,20,20,22,24,24,24,26,28,28}. The corresponding OEIS-entry is \url{http://oeis.org/A007843} (which has $1$ as first element (index $0$), instead of $0$ as we have it, and which we regard as more appropriate).
\end{examp}

\begin{lem}[\cite{WenpengLiu2002Smarandache}]\label{lem:boundsSp}
  For $k \in \NN$ holds $k + 1 \le S_2(k) = k + O(\log k)$.
\end{lem}
We give an independent proof for the lower bound in Lemma \ref{lem:lbS'}, while we sharpen the upper bound in Theorem \ref{thm:S2upper}. For more number-theoretic properties of $S_2$ see \cite{Wenpeng2004SmarandacheProblems}. To understand the plateaus of $S_2$, we need the \emph{ruler function}:
\begin{defi}\label{def:rseq}
  Let $\ruler_n := \ord_2(2 n) \in \NN$ for $n \in \NN$.
\end{defi}

\begin{examp}\label{exp:rseq}
  The first $30$ elements of $\ruler_n$ are \mathlist{1, 2, 1, 3, 1, 2, 1, 4, 1, 2, 1, 3, 1, 2, 1, 5, 1, 2, 1, 3, 1, 2, 1, 4, 1, 2, 1, 3, 1, 2} (\url{http://oeis.org/A001511}).
\end{examp}

The plateaus of $S_2$ are given by the ruler function: in Lemma \ref{lem:S_pseq} we determined the number of repetitions of values $v \in \NN$ as $\ord_2(v)$, while for the plateaus we skip zero-repetitions, which happen at each odd number, and thus for the associated index $n$ we have $n = \frac v2$ for even $v$, and the number of repetitions is $\ord_2(v) = \ord_2(2 n)$; we obtain
\begin{lem}\label{lem:smaruler}
  $\platop (S_2(k))_{k \in \NN} = (\ruler_n)_{n \in \NN}$.
\end{lem}

\subsection{Meta-Fibonacci sequences}
\label{sec:smaaltrec}

Started by \cite[Page 145]{Hofstadter1979GEB}, various nested recursions for integer sequences have been studied. Often the focus in this field of ``meta-Fibonacci sequences'' is on ``chaotic behaviour'', but we consider here only a well-behaved case (but in detail):
\begin{defi}\label{def:metafib}
  In \cite{RuskeyDeugau2009MetaFibonacci} the sequence $a_2: \NNZ \ra \NNZ$\footnote{hiding two parameters $d \in \NN$, $s \in \ZZ$ used in \cite{RuskeyDeugau2009MetaFibonacci}, which are $d=2$, $s=0$ in our case}, has been defined recursively via
  \begin{displaymath}
    a_2(k) = a_2(k - a_2(k-1)) + a_2(k - 1 - a_2(k-2)),
  \end{displaymath}
  while $a_2(k) := k$ for $k \in \set{0,1}$.
\end{defi}

The sequence $a_2$ was introduced in \cite{Conolly1989MetaFibonacci} as $F: \NN \ra \NNZ$, with $F(k) = k-1$ for $k \in \set{1,2}$ and the same recursion law, which yields $F(k) = a_2(k-1)$ for $k \in \NN$. Furthermore, using $F'(1) = F'(2) = 1$ as initial conditions does not change anything else, and this sequence is the OEIS entry \url{http://oeis.org/A046699}.

\begin{examp}\label{exp:metafib}
  Numerical values for $a_2(k)$ and $k \in \tb{0}{27}$: \mathlist{0, 1, 2, 2, 3, 4, 4, 4, 5, 6, 6, 7, 8, 8, 8, 8, 9, 10, 10, 11, 12, 12, 12, 13, 14, 14, 15}.  The first five recursive computations:
    \begin{enumerate}
    \item $a_2(2) = a_2(2 - a_1(1)) + a_2(1 - a_2(0)) = a_2(2 - 1) + a_2(1 - 0) = a_2(1) + a_2(1) = 1 + 1 = 2$.
    \item $a_2(3) = a_2(3 - a_2(2)) + a_2(2 - a_2(1)) = a_2(3 - 2) + a_2(2 - 1) = a_2(1) + a_2(1) = 1 + 1 = 2$.
    \item $a_2(4) = a_2(4 - a_2(3)) + a_2(3 - a_2(2)) = a_2(4 - 2) + a_2(3 - 2) = a_2(2) + a_2(1) = 2 + 1 = 3$.
    \item $a_2(5) = a_2(5 - a_2(4)) + a_2(4 - a_2(3)) = a_2(5 - 3) + a_2(4 - 2) = a_2(2) + a_2(2) = 2 + 2 = 4$.
    \item $a_2(6) = a_2(6 - a_2(5)) + a_2(5 - a_2(4)) = a_2(6 - 4) + a_2(5 - 3) = a_2(2) + a_2(2) = 2 + 2 = 4$.
    \end{enumerate}
\end{examp}

It is shown (in our notation):
\begin{lem}[\cite{Conolly1989MetaFibonacci}]\label{lem:specreca2}
  For $k \in \NN$ and $p := \floor{\log_2(k+1)}$: $a_2(k) = 2^{p-1} + a_2(k+1-2^p)$.
\end{lem}
Lemma \ref{lem:specreca2} yields a fast computation of $a_2(k)$. \cite[Corollary 2.9, Equation (1)]{JacksonRuskey2006MetaFibonacci} determines the plateau sizes:
\begin{lem}[\cite{JacksonRuskey2006MetaFibonacci}]\label{lem:pota2g}
  $a_2$ is a $1$-Delta sequence with $\platop (a_2(k))_{k \in \NN} = \ruler$.
\end{lem}

We can now show $a_2 = \frac 12 S_2$, which has been conjectured on the OEIS (\url{http://oeis.org/A007843}, by Michel Marcus):
\begin{thm}\label{thm:Spad}
  $\fa\, k \in \NNZ: S_2(k) = 2 \cdot a_2(k)$.
\end{thm}
\begin{prf}
By Lemma \ref{lem:smaruler} and Lemma \ref{lem:pota2g}, together with $S_2(0) = a_2(0) = 0$. \Qed
\end{prf}

\section{Recursions for Smarandache primitive numbers}
\label{sec:recSd}

In Subsection \ref{sec:courseofvalues} we introduce the sequence $S'_2$ via a recursive process, which directly ties into our main application in Theorem \ref{thm:smalbfccon}, for constructing unsatisfiable hitting clause-sets with many full clauses. This recursive definition uses an index, which is studied in Subsection \ref{sec:simprecd}. The central helper function is the ``slack'', studied in Subsection \ref{sec:slack}. We then prove a meta-Fibonacci recursion in Theorem \ref{thm:nestrecS'}, and obtain $S'_2 = S_2$ in Theorem \ref{thm:smarecfac}.

\subsection{A simple course-of-values recursion}
\label{sec:courseofvalues}

\begin{defi}\label{def:recsmaran}
  For $k \in \NNZ$ let
  \begin{enumerate}
  \item $S'_2(0) := 0$, $S'_2(1) := 2$; and for $k \ge 2$:
  \item $S'_2(k) := 2 \cdot (k-i+1)$ for the minimal $i \in \tb {1}{k-1}$ with $k-i+1 \le S'_2(i)$.
  \end{enumerate}
\end{defi}
Note that the recursion step is well-defined (the $i$ exists), since for $i=k-1$ holds $k-i+1 = 2$, and $S'_2(k-1) = 2$ for $k=2$, while for $k \ge 3$ holds $S'_2(k-1) = 2 \cdot ((k-1)-i'+1) \ge 2 \cdot ((k-1) - ((k-1)-1)+1) = 4$. The condition ``$k-i+1 \le S'_2(i)$'' is equivalent to $k+1 \le i + S'_2(i)$. Some simple properties are that $S'_2(k)$ is divisible by $2$, $S'_2(k) \ge 2$ for $k \ge 1$, and $S'_2(2) = 4$ and $S'_2(k) \ge 4$ for $k \ge 2$.

\begin{examp}\label{exp:S'd}
  The computations for $S'_2(k)$ for $1 \le k \le 10$:
  \begin{enumerate}
  \item[$1 \ra$] $\bmm{2}$ by recursion basis
  \item[$2 \ra$] $2 \cdot (2-1+1) = \bmm{4}$; $1 + 2 = 3 \ge 3$ ($i=1$)
  \item[$3 \ra$] $2 \cdot (3-2+1) = \bmm{4}$; $2 + 4 = 6 \ge 4$ ($i=2$)
  \item[$4 \ra$] $2 \cdot (4-2+1) = \bmm{6}$; $2 + 4 = 6 \ge 5$ ($i=2$)
  \item[$5 \ra$] $2 \cdot (5-2+1) = \bmm{8}$; $2 + 4 = 6 \ge 6$ ($i=2$)
  \item[$6 \ra$] $2 \cdot (6-3+1) = \bmm{8}$; $3 + 4 = 7 \ge 7$ ($i=3$)
  \item[$7 \ra$] $2 \cdot (7-4+1) = \bmm{8}$; $4 + 6 = 10 \ge 8$ ($i=4$)
  \item[$8 \ra$] $2 \cdot (8-4+1) = \bmm{10}$; $4 + 6 = 10 \ge 9$ ($i=4$)
  \item[$9 \ra$] $2 \cdot (9-4+1) = \bmm{12}$; $4 + 6 = 10 \ge 10$ ($i=4$)
  \item[$10 \ra$] $2 \cdot (10-5+1) = \bmm{12}$; $5 + 8 = 13 \ge 11$ ($i=5$).
  \end{enumerate}
\end{examp}

\subsection{Analysing the index}
\label{sec:simprecd}

\begin{defi}\label{def:Si}
  For $k \ge 0$ let $\bmm{\ismar(k)} := k+1 - \frac{S'_2(k)}2 \in \NN$.
\end{defi}
Simple properties (for all $k \ge 0$):
\begin{enumerate}
\item $S'_2(k) = 2 \cdot (k -\ismar(k) + 1)$.
\item $\ismar(0) = \ismar(1) = \ismar(2) = 1$.
\item $\Delta \ismar(k) = 0 \Lra \Delta S'_2(k) = 2$ and $\Delta \ismar(k) = 1 \Lra \Delta S'_2(k) = 0$.
\end{enumerate}

\begin{examp}\label{exp:Si}
  Numerical values of $\ismar(k)$ for $k \in \tb 0{25}$ are, together with $S'_2(k)$, $S'_2(\ismar(k))$, and the sum of first and third row minus $k+1$, which is denoted below by ``$\ssmar(k)$'':

  \noindent$1,1,1,2,2,2,3,4,4,4,5,5,5,6,7,8,8,8,9,9,9,10,11,11,11,12$.\\
  $0,2,4,4,6,8,8,8,10,12,12,14,16,16,16,16,18,20,20,22,24,24,24,26,28,28$\\
  $2,2,2,4,4,4,4,6,6,6,8,8,8,8,8,10,10,10,12,12,12,12,14,14,14,16$\\
  $2,1,0,2,1,0,0,2,1,0,2,1,0,0,0,2,1,0,2,1,0,0,2,1,0,2$.
\end{examp}

An alternative characterisation of $\ismar(k)$:
\begin{lem}\label{lem:altcharacid}
  For $k \ge 0$: $\ismar(k)$ is the minimal $i \in \NNZ$ with $i + S'_2(i) \ge k+1$.
\end{lem}
\begin{prf}
The assertion follows by what has already been said above, plus the consideration of the corner cases: $0 + S'_2(0) = 0 < k+1$ for all $k \ge 0$, while $1 + S'_2(1) = 3 \ge k+1$ for $k \le 2$. \Qed
\end{prf}

We obtain a method to prove lower bounds for $S'_2(k)$:
\begin{corol}\label{cor:altcharacid}
  For $k, i \in \NNZ$ with $S'_2(i) \ge k-i+1$ holds $S'_2(k) \ge 2 (k-i+1)$.
\end{corol}

$\ismar(k)$ grows in steps of $+1$, while $S'_2(k)$ grows in steps of $+2$:
\begin{lem}\label{lem:DeltaS'i}
  $\Delta S'_2(k) \in \set{0,2}$ and $\Delta \ismar(k) \in \set{0,1}$ for all $k \in \NNZ$.
\end{lem}
\begin{prf}
 Proof via (simultaneous) induction on $k$: The assertions hold for $k \le 1$, and so consider $k \ge 2$. Now $\ismar(k)$ is the minimal $i \in \tb {1}{k-1}$ with $k+1 \le i+S'_2(i)$, and due to $\Delta S'_2(i) \ge 0$ for all $i < k$ it follows $\Delta \ismar(k) \in \set{0,1}$. \Qed
\end{prf}

We obtain a simple upper bound on $\ismar$:
\begin{corol}\label{cor:simbid}
  For $k \ge 1$ holds $\ismar(k) \le k$ and for $k \ge 2$ holds $\ismar(k) \le k-1$
\end{corol}

\subsection{The ``slack''}
\label{sec:slack}

An important helper function is the ``slack'' $\ssmar(k)$:
\begin{defi}\label{def:gapid}
  For $k \in \NNZ$ let $\bmm{\ssmar(k)} := (\ismar(k) + S'_2(\ismar(k))) - (k+1) \in \NNZ$.
\end{defi}
So $\ssmar(0) = (1 + 2) - (0 + 1) = 2$ and $\ssmar(1) = (1 + 2) - (1 + 1) = 1$. Directly from the definition follows:

\begin{lem}\label{lem:S'id}
  For $k \ge 0$ holds $S'_2(\ismar(k)) = \frac 12 S'_2(k) + \ssmar(k)$.
\end{lem}

We can characterise the cases $\Delta \ismar(k) = 1$ as the ``slackless'' $k$'s:
\begin{lem}\label{lem:eqidD}
  For $k \ge 0$:
  \begin{enumerate}
  \item $\Delta \ismar(k) = 1 \Lra \ssmar(k) = 0 \Lra \Delta S'_2(k) = 0$.
  \item $\Delta \ismar(k) = 0 \Lra \ssmar(k) \ge 1 \Lra \Delta S'_2(k) = 2$.
  \end{enumerate}
\end{lem}
\begin{prf}
If $\ssmar(k) \ge 1$, then $\Delta \ismar(k) = 0$ by Lemma \ref{lem:altcharacid}, while for $\ssmar(k) = 0$ we get $\Delta \ismar(k) \ge 1$. \Qed
\end{prf}

Thus the slack determines the growth of $S'_2$:
\begin{corol}\label{cor:slackvsD}
  For $k \ge 0$ holds $\Delta S'_2(k) = 2 \cdot \min(\ssmar(k),1)$.
\end{corol}

And plateaus of the slack happen only for slack zero, and from such a plateau the slack jumps to $2$, and then is stepwise again decremented to zero:
\begin{corol}\label{cor:behavslack}
  For $k \ge 0$ holds:
  \begin{enumerate}
  \item If $\ssmar(k) > 0$, then $\ssmar(k+1) = \ssmar(k) - 1$.
  \item If $\ssmar(k) = 0$, then $\ssmar(k+1) \in \set{0,2}$.
  \end{enumerate}
\end{corol}

\subsection{A meta-Fibonacci recursion}
\label{sec:S'meta}

We are ready to prove an interesting nested recursion for $S'_2$. First a combinatorial lemma, just exploiting the fact that the shape of the slack repeats the following pattern (Corollary \ref{cor:behavslack}): a plateau of zeros, followed by a jump to $2$ and a stepwise decrement to $0$ again (where right at $k=0$ we start with $\ssmar(0) = 2$):
\begin{lem}\label{lem:sumslack}
  For $k \ge 2$ holds $\sum_{i=1}^2 \ssmar(k-i) = \sum_{i=1}^2 i \cdot \min(1, \ssmar(k-i))$.
\end{lem}
\begin{prf}
There are $0 \le p \le 2$ and $1 \le q \le 3$ such that the left-hand side is
\begin{displaymath}
  p + (p-1) + \dots + 1 + 0 + \dots + 0 + 2 + (2-1) + \dots + q;
\end{displaymath}
for $p = 0$ the initial part is empty, for $q=3$ the final part is empty. Let $r \ge 0$ be the number of zeros; so $r = 0$ iff $p=2$ (and then also $q=3$). We have $p + r + (2-q+1) = 2$, i.e., $p + r + 1 = q$. Now the right-hand side is
\begin{displaymath}
  1 + 2 + \dots + p + 0 + \dots + 0 + q + (q+1) + \dots + 2,
\end{displaymath}
and we see that both sides are equal. \Qed
\end{prf}

\begin{thm}\label{thm:nestrecS'}
  For $k \ge 2$ holds
  \begin{displaymath}
    S'_2(k) = \sum_{i=1}^2 S'_2(\ismar(k-i))
  \end{displaymath}
  (note that by Lemma \ref{cor:simbid} holds $\ismar(k-i) < k$).
\end{thm}
\begin{prf}
By Lemma \ref{lem:S'id} and Lemma \ref{lem:sumslack} holds
\begin{multline*}
  \sum_{i=1}^2 S'_2(\ismar(k-i)) = (\sum_{i=1}^2 \ssmar(k-i)) + S'_2(k) - \frac 12 \sum_{i=1}^2 (S'_2(k) - S'_2(k-i)) =\\
  S'_2(k) + (\sum_{i=1}^2 i \cdot \min(1, \ssmar(k-i))) - \frac 12 \sum_{i=1}^2 \sum_{j=0}^{i-1} \Delta S'_2(k+i-j),
\end{multline*}
where now by Corollary \ref{cor:slackvsD} holds $\sum_{i=1}^2 \sum_{j=0}^{i-1} \Delta S'_2(k+i-j) = (\Delta S'_2(k-1)) + (\Delta S'_2(k-2) + \Delta S'_2(k-1)) = \sum_{i=1}^2 i \cdot \Delta S'_2(k-1) = 2 \sum_{i=1}^2 i \cdot \min(1,\ssmar(k))$, which completes the proof. \Qed
\end{prf}

Now we see that $S'_2$ is basically the same as $a_2$ (recall Subsection \ref{sec:smaaltrec}):
\begin{corol}\label{cor:metafib}
  $\fa\, k \in \NNZ: S_2'(k) = 2 \cdot a_2(k)$.
\end{corol}
\begin{prf}
For the purpose of the proof let $a_2(k) := \frac 12 S_2'(k)$ for $k \in \NNZ$. So we get $a_2(k) = k$ for $k \in \set{0,1}$, while $\ismar(k) = k+1 - a_2(k)$, and thus for $k \ge 2$:
\begin{multline*}
  a_2(k) = \frac 12 S'_2(k) = \frac 12 \sum_{i=1}^2 S'_2(\ismar(k-i)) = \sum_{i=1}^2 a_2(\ismar(k-i)) =\\
  \sum_{i=1}^2 a_2(k-i+1 - a_2(k-i)),
\end{multline*}
and so the assertion follows by the equations of Definition \ref{def:metafib}. \Qed
\end{prf}

We obtain the main result of this section:
\begin{thm}\label{thm:smarecfac}
  $S'_2 = S_2$ (recall Definition \ref{def:smafac}).
\end{thm}
\begin{prf}
By Corollary \ref{cor:metafib} and Theorem \ref{thm:Spad}. \Qed
\end{prf}

\section{On the number of full clauses}
\label{sec:onfc}

First we review full subsumption resolution, $C \cup \set{v}, C \cup \set{\ol{v}} \leadsto C$, and its inversion, called ``extension'' in Section \ref{sec:fullsr}, where some care is needed, since we need complete control.
From a clause-set $F$ with ``many'' full clauses we can produce further clause-sets with ``many'' full clauses by full subsumption extension done in parallel, and this process of ``full expansion'' is presented in Definition \ref{def:msext}. The recursive computation of $S_2$ via Definition \ref{def:recsmaran} captures maximisation for this process, and so we can show in Theorem \ref{thm:smalbfccon}, that we can construct examples of unsatisfiable hitting clause-sets $F_k$ of deficiency $k$ and with $S_2(k)$ many full clauses. It follows that $S_2$ yields a lower bound on $\maxsmarh$ (Conjecture \ref{con:S2exact} says this lower bound is actually an equality).

\subsection{Full subsumption resolution}
\label{sec:fullsr}

As studied in \cite[Section 6]{KullmannZhao2010Extremal} in some detail:
\begin{defi}[\cite{KullmannZhao2010Extremal}]\label{def:fullsubres}
  A \textbf{full subsumption resolution} for $F \in \Cls$ can be performed, if there is a clause $C \notin F$ with $C \cup \set{v}, C \cup \set{\ol{v}} \in F$ for some variable $v$, and replaces the two clauses $C \cup \set{v}, C \cup \set{\ol{v}}$ by the single clause $C$. For the \textbf{strict} form, there must exist a third clause $D \in F \sm \set{C \cup \set{v}, C \cup \set{\ol{v}}}$ with $v \in \var(D)$, while for the \textbf{non-strict} form there must NOT exist such a third clause.
\end{defi}
If $F'$ is obtained from $F$ by one full subsumption resolution, then $c(F') = c(F) -1$; we have the strict form iff $n(F') = n(F)$, or, equivalently, $\delta(F') = \delta(F) - 1$, while we have the non-strict form iff $n(F') = n(F)- 1$, or, equivalently, $\delta(F') = \delta(F)$. A very old transformation of a CNF (DNF) into an equivalent one uses the inverse of full subsumption resolution\footnote{Boole introduced in \cite{Boole1854Gesetzeb}, Chapter 5, Proposition II, the general ``expansion'' $f(v,\vec{x}) = (f(0,\vec{x}) \wedge \ol{v}) \vee (f(1,\vec{x}) \wedge v)$ for boolean functions $f$, where for our application $f(v,\vec{x}) \approx C$. This was taken up by \cite{Shannon1940Schalter}, and is often referred to as ``Shannon expansion''.}:
\begin{defi}[\cite{KullmannZhao2010Extremal}]\label{def:fullsubext}
  A \textbf{full subsumption extension} for $F \in \Cls$ and a clause $C \in F$ can be performed, if there is a variable $v \in \Va \sm \var(C)$ with $C \cup \set{v}, C \cup \set{\ol{v}} \notin F$, and replaces the single clause $C$ by the two clauses $C \cup \set{v}, C \cup \set{\ol{v}}$. For the \textbf{strict} form we have $v \in \var(F)$, while for the \textbf{non-strict} form we have $v \notin \var(F)$.
\end{defi}
If we consider $F \in \Musat$ and $C \in F$, then we can always perform a non-strict full subsumption extension, while we can perform the strict form iff $C$ is not full. If we denote the result by $F'$, then for $F \in \Uclash$ we have again $F' \in \Uclash$, but for general $F \in \Musat$ we might have $F' \notin \Musat$; see \cite[Lemma 6.5]{KullmannZhao2010Extremal} for an exact characterisation.

\subsection{Full expansions}
\label{sec:exp}

We now perform full subsumption extensions in parallel to $m$ full clauses of $F$, first using a non-strict extension, and then reusing the extension variable via strict extensions:
\begin{defi}\label{def:msext}
  For $F \in \Cls$ and $m \in \NN$, where $\nfc(F) \ge m$, a \textbf{full $m$-expansion of $F$} is some $G \in \Cls$ obtained by
  \begin{enumerate}
  \item choosing some $F' \sse F \cap A(\var(F))$ with $c(F') = m$,
  \item choosing some $v \in \Va \sm \var(F)$ (the \textbf{extension variable}),
  \item and replacing the clauses $C \in F'$ in $F$ by their full subsumption extension with $v$ (recall Definition \ref{def:fullsubext}).
  \end{enumerate}
\end{defi}

The choice of $v$ in Definition \ref{def:msext} is irrelevant, while the choice of $F'$ might have an influence on further properties of $G$, but is irrelevant for our uses. The following basic properties all follow directly from the definition:
\begin{lem}\label{lem:propmsext}
  Consider the situation of Definition \ref{def:msext}.
  \begin{enumerate}
  \item There is always a full $m$-expansion $G$ (unique for any fixed $F'$, $v$).
  \item If $F \in \Uclash$, then $G \in \Uclash$.
  \item $n(G) = n(F) + 1$, $c(G) = c(F) + m$.
  \item $\delta(G) = \delta(F) + m-1$.
  \item $\nfc(G) = 2 \cdot m$.
  \end{enumerate}
\end{lem}

We turn to the construction of unsatisfiable hitting clause-sets with many full clauses (for a given deficiency):
\begin{thm}\label{thm:smalbfccon}
  For $k \in \NN$ we recursively construct $F_k \in \Uclashi{\delta=k}$ as follows:
  \begin{enumerate}
  \item $F_1 := \set{\set{1},\set{-1}}$.
  \item For $k \ge 2$ let $F_k$ be a full $a_2(k)$-expansion of $F_{\ismar(k)}$.
  \end{enumerate}
  Then we have $\nfc(F_k) = S_2(k)$. Thus $\fa\, k \in \NN : S_2(k) \le \maxsmarh(k)$.
\end{thm}
\begin{prf}
If the construction is well-defined, then we get $\nfc(F_k) = 2 \cdot a_2(k) = S_2(k)$ and $\delta(F_k) = \delta(F_{\ismar(k)}) + a_2(k)-1 = \ismar(k) + a_2(k)-1 = k$ for $k \ge 2$ by Lemma \ref{lem:propmsext} (using Theorem \ref{thm:smarecfac} freely), while these two properties hold trivially for $k=1$.

It remains to show that $1 \le \ismar(k) \le k-1$ and $a_2(k) \le \nfc(F_{\ismar(k)})$ for $k \ge 2$. The first statement follows by Corollary \ref{cor:simbid}, while the second statement follows by Lemma \ref{lem:altcharacid}. \Qed
\end{prf}

\section{Applications}
\label{sec:applications}

We start by sharpening the upper bound from Lemma \ref{lem:boundsSp}:
\begin{thm}\label{thm:S2upper}
  For $k \in \NN$ holds $S_2(k) \le \nonmer(k) \le k + 1 + \floor{\log_2(k)}$.
\end{thm}
\begin{prf}
By Theorem \ref{thm:smalbfccon} and Theorem \ref{thm:uppernM}. \Qed
\end{prf}

We can also provide an independent proof of the lower bound of Lemma \ref{lem:boundsSp}:
\begin{lem}\label{lem:lbS'}
  For $k \in \NN$ holds $S_2(k) \ge k+1$.
\end{lem}
\begin{prf}
We prove the assertion by induction. For $k=1$ we have $S_2(1) = 2$, so consider $k \ge 2$. We use Corollary \ref{cor:altcharacid}, and so we need $i \in \NN$ with $k + 1 \le 2 (k - i + 1)$, i.e., $i \le \frac{k+1}{2}$. So we choose $i := \floor{\frac{k+1}{2}} \in \NN$. We have $i < k$, and so we can apply the induction hypothesis to $i$: $i + S_2(i) = \floor{\frac{k+1}2} + S_2(\floor{\frac{k+1}2}) \ge \floor{\frac{k+1}2} + \floor{\frac{k+1}2} + 1 = 2 \floor{\frac{k+1}2} + 1 > 2 (\frac{k+1}2 - 1) + 1 = k$, and thus $i + S_2(i) \ge k+1$. \Qed
\end{prf}

When upper and lower bound coincide, then we know all four fundamental quantities; first we name the sets of deficiencies (recall Theorems \ref{thm:uppernM}, \ref{thm:uppernM1}):
\begin{defi}\label{def:equplb}
  $\bmm{\Esnm} := \set{k \in \NN : S_2(k) = \nonmer(k)}$, $\bmm{\Esnm_1} := \set{k \in \NN : S_2(k) = \nonmer_1(k)}$.
\end{defi}
By $S_2 \le \minnonmer \le \nonmer_1 \le \nonmer$ we get $\Esnm \sse \Esnm_1$ and:
\begin{thm}\label{thm:meanesnm}
  For $k \in \Esnm_1$ holds $S_2(k) = \maxsmarh(k) = \maxsmar(k) = \minnonmerh(k) = \minnonmer(k) = \nonmer_1(k)$.
\end{thm}

We prove now that the special deficiencies $2^n - n, 2^n - n - 1$ ($n \ge 1$; note $\delta(A_n) = 2^n-n$) considered in \cite[Lemmas 12.10, 12.11]{KullmannZhao2010Extremal}, where we have shown that for them the four fundamental quantities coincide, are indeed in $\Esnm$, and that furthermore the special deficiencies $2^n - n + 1$ ($n \ge 3$), where $\nonmer_1$ differs from $\nonmer$, are in $\Esnm_1$:
\begin{lem}\label{lem:specdef}
  Consider $n \in \NN$.
  \begin{enumerate}
  \item\label{lem:specdef0} $S_2(2^n - n) = 2^n$, and for $k \in \NNZ$ holds $S_2(k) = 2^n \Lra 2^n - n \le k \le 2^n - 1$.
  \item\label{lem:specdef1} $2^n - n \in \Esnm$, while $2^n - n + 1, \dots, 2^n - 1 \notin \Esnm$.
  \item\label{lem:specdef2} Assume $n \ge 2$ now. Then $2^n - n - 1 \in \Esnm$ with $S_2(2^n - n - 1) = 2^n - 2$.
  \item\label{lem:specdef3} For $n \ge 3$ holds $2^n - n + 1 \in \Esnm_1$.
  \end{enumerate}
\end{lem}
\begin{prf}
By \cite[Corollary 7.24]{KullmannZhao2010Extremal} we have $\nonmer(2^n-n) = 2^n$, while $\nonmer(2^n - n - 1) = 2^n - 2$ (remember that the jumps for $\nonmer$ happens at the deficiencies $2^n - n$). Thus $S_2(2^n-n) \le 2^n$ and $S_2(2^n-n-1) \le 2^n - 2$. Since for the value $2^n$ the sequence $S_2$ has a plateau of length $n$ (Lemma \ref{lem:S_pseq}), while $\nonmer$ is strictly increasing, for Parts \ref{lem:specdef0}, \ref{lem:specdef1}, \ref{lem:specdef2} it remains to show $S_2(2^n-n) \ge 2^n$. We show this by induction: For $n=1$ we have $S_2(1) = 2 = 2^1$, while for $n \ge 2$ by induction hypothesis we have $(2^n - n) - (2^{n-1} - (n-1)) +1 = 2^{n-1} \le S_2(2^{n-1} - (n-1))$, thus by Corollary \ref{cor:altcharacid} $S_2(2^n-n) \ge 2 \cdot 2^{n-1} = 2^n$. Finally, for Part \ref{lem:specdef3} we note $S_2(2^n - n + 1) = S_2(n) = 2^n$ by Part \ref{lem:specdef0}, while $\nonmer_1(k)$ differs from $\nonmer(k)$ exactly at the positions $k = 2^n-n+1$ for $n \ge 3$, where then $\nonmer_1(k) = \nonmer(k) - 1 = 2^n$ (\cite[Theorem 14.7]{KullmannZhao2010Extremal}). \Qed
\end{prf}

So the lower bound of Lemma \ref{lem:lbS'} is sharp for infinitely many deficiencies:
\begin{corol}\label{cor:lowbdatt}
  We have $S_2(k) = k+1$ for all $k = 2^n-1$, $n \in \NN$.
\end{corol}

\section{Initial values of the four fundamental quantities}
\label{sec:initvalues}

The task of this penultimate section is to prove the values in Table \ref{tab:valuesmaxnfc} (in Theorem \ref{thm:13values}; of course, only the four fundamental quantities are open).

\begin{table}[h]
\large\setlength{\tabcolsep}{6pt}
  \centering
  \begin{tabular}{c||c|*{3}{c}|*{7}{c}|*{2}{c}}
    $k$ & 1 & 2 & \bmm{3} & 4 & 5 & 6 & \bmm{7} & \bmm{8} & 9 & \bmm{10} & 11 & 12 & 13\\
    \hline
    \hline
    $\nonmer(k)$ & 2 & 4 & \bmm{5} & 6 & 8 & 9 & \bmm{10} & \bmm{11} & 12 & \bmm{13} & 14 & 16 & 17\\
    \hline
    $\nonmer_1(k)$ & 2 & 4 & \bmm{5} & 6 & 8 & 8 & \bmm{10} & \bmm{11} & 12 & \bmm{13} & 14 & 16 & 16\\
    \hline
    $\minnonmer(k)$ & 2 & 4 & \bmm{5} & 6 & 8 & 8 & \bmm{10} & \bmm{11} & 12 & \bmm{13} & 14 & 16 & 16\\
    \hline
    $\minnonmerh(k)$ & 2 & 4 & \bmm{5} & 6 & 8 & 8 & \bmm{10} & \bmm{11} & 12 & \bmm{13} & 14 & 16 & 16\\
    \hline
    $\maxsmar(k)$ & 2 & 4 & \bmm{4} & 6 & 8 & 8 & \bmm{9} & \bmm{10} & 12 & \bmm{12} & 14 & 16 & 16\\
    \hline
    $\maxsmarh(k)$ & 2 & 4 & \bmm{4} & 6 & 8 & 8 & \bmm{8} & \bmm{10} & 12 & \bmm{12} & 14 & 16 & 16\\
    \hline
    $S_2(k)$ & 2 & 4 & \bmm{4} & 6 & 8 & 8 & \bmm{8} & \bmm{10} & 12 & \bmm{12} & 14 & 16 & 16
  \end{tabular}
  \caption{Values for the fundamental quantities for $1 \le k \le 13$; in bold the columns not in $\Esnm_1$, while the vertical bars are left of the special deficiencies $2^n - n$, $n \ge 2$.}
  \label{tab:valuesmaxnfc}
\end{table}

Strengthening \cite[Corollary 12.13]{KullmannZhao2010Extremal}, first we establish properties of $F \in \Musat$ such that the number of full clauses equals the min-var-degree, i.e., there is a variable which occurs only in the full clauses. We use $\varmvd(F) := \set{v \in \var(F) : \vdeg_F(v) = \minvdeg(F)}$ for $F \in \Cls$ with $n(F) > 0$ (the set of variables with minimal degree). Furthermore we use $\dpi{v}(F)$ (``DP-reduction'', also called ``variable elimination''; see \cite{KullmannZhao2012ConfluenceJ} for more on this important operation) for $F \in \Cls$ and $v \in \var(F)$ for the result of replacing the clauses containing variable $v$ by their ``resolvents'' on $v$, which for clauses $C, D \in F$ with $v \in C$, $\ol{v} \in D$ is $(C \sm \set{v}) \cup (D \sm \set{\ol{v}})$, and is only defined in case $C, D$ do not have other clashes. Indeed the special use in Lemma \ref{lem:evenfc} yields the inverse of the expansion process from Definition \ref{def:msext}.

\begin{lem}\label{lem:evenfc}
  Consider $F \in \Musat$ with $\nfc(F) = \minvdeg(F)$ (and thus $n(F) > 0$).
  \begin{enumerate}
  \item\label{lem:evenfc0} $\varmvd(F)$ is the set of all $v \in \var(F)$ which occur only in full clauses of $F$.
  \item\label{lem:evenfc1} $\nfc(F)$ is even.
  \item\label{lem:evenfc3} For $v \in \varmvd(F)$ and $F' := \dpi{v}(F)$ we have $F' \in \Musati{\delta=\delta(F)-\frac {\nfc(F)}2 +1}$.
  \item\label{lem:evenfc2} $\nfc(F) \le 2 \cdot \maxsmar(\delta(F)-\frac {\nfc(F)}2 +1)$.
  \end{enumerate}
\end{lem}
\begin{prf}
Consider $v \in \var(F)$ with $\vdeg_F(v) = \minvdeg(F)$. The occurrences of $v$ are now exactly in the full clauses of $F$ (Part \ref{lem:evenfc0}). Every full clauses must be resolvable on $v$, and thus the full clauses of $F$ can be partitioned into pairs $\set{v} \addcup C, \set{\ol{v}} \addcup C$ for $\frac{\nfc(F)}2$ many clauses $C$. This shows Part \ref{lem:evenfc1}. Parts \ref{lem:evenfc3}, \ref{lem:evenfc2} now follow by considering $F' := \dpi{v}(F)$: $F'$ is obtained by replacing the full clauses of $F$ by the clauses $C$ (i.e., performs a full subsumption resolution, which are all strict except of the last one, which is non-strict). The new clauses $C$ are full in $F'$ (though there might be other full clauses in $F'$). Obviously $F' \in \Musat$ and $\delta(F') = \delta(F) - \frac{\nfc(F)}2 + 1$. \Qed
\end{prf}

For deficiency $k=7$ we have the first case of $\maxsmarh(k) < \maxsmar(k)$:
\begin{lem}\label{lem:nfcdef7}
  $\maxsmar(7) = 9 = \nonmer(7)-1$, while $\maxsmarh(7) = 8 = S_2(7)$.
\end{lem}
\begin{prf}
By $S_2(7) = 8$ we have $\maxsmarh(7) \ge 8$. By Lemma \ref{lem:evenfc}, Part \ref{lem:evenfc2} and by $\maxsmar(3) = 4$ the assumption of $\maxsmar(7) = 10 = \nonmer(7)$ yields the contradiction $10 \le 2 \maxsmar(7 - 5 + 1) = 2 \cdot 4 = 8$, and thus $\maxsmar(7) \le 9$. By Lemma \ref{lem:evenfulluhit} we obtain $\maxsmarh(7) = 8$. A clause-set $F \in \Musati{\delta=7}$ with $\nfc(F) = 9$ (and $n(F) = 4$) is given by the following variable-clause-matrix (the clauses are the columns):
\setcounter{MaxMatrixCols}{11}
\begin{displaymath}
  \begin{pmatrix}
    - & - & + & + & - & - & + & - & - & + & 0\\
    + & + & - & - & - & - & + & - & + & - & 0\\
    + & - & + & - & + & - & 0 & + & + & + & -\\
    + & + & + & + & + & + & 0 & - & - & - & -
  \end{pmatrix}
\end{displaymath}
Let the variables be $1, \dots, 4$, as indices of the rows. Now setting variable $4$ to \texttt{false} yields $A_3$, where one non-strict subsumption resolution has been performed, while setting variable $4$ to \texttt{true} followed by unit-clause propagation of $\set{-3}$ yields $A_2$. So both instantiations yield minimally unsatisfiable clause-sets, whence by \cite[Lemma 3.15, Part 2]{KullmannZhao2010Extremal} $F \in \Musat$.\footnote{\cite[Lemma 3.15]{KullmannZhao2010Extremal} contains a technical correction over \cite[Lemma 1]{KullmannZhao2011Bounds}.} \Qed
\end{prf}

We are ready to prove the final main result of this \Schrift{}:
\begin{thm}\label{thm:13values}
  Table \ref{tab:valuesmaxnfc} is correct.
\end{thm}
\begin{prf}
The values for $1 \le k \le 6$ have been determined in \cite[Section 14]{KullmannZhao2010Extremal}. We observe that $1,2,4,5,6,9,11,12,13 \in \Esnm_1$, and thus by Theorem \ref{thm:meanesnm} nothing is to be done for these values, and only the deficiencies $7, 8, 10$ remain.

By Lemma \ref{lem:evenfc}, Part \ref{lem:evenfc1}, we get that $\maxsmarh(8) = \maxsmar(8) = 10$ (since $\nonmer(8) = 11$ is odd), and also $\maxsmarh(10) = \maxsmar(10) = 12$. By Lemma \ref{lem:nfcdef7} it remains to provide unsatisfiable hitting clause-sets witnessing $\minnonmerh(7) = 10$, $\minnonmerh(8) = 11$ and $\minnonmerh(10) = 13$. For deficiency $7$ consider
    \begin{displaymath}
      F_7 :=
      \begin{pmatrix}
        0 & + & - & + & - & + & - & - & + & - & +\\
        0 & - & + & + & - & - & + & - & - & + & +\\
        - & + & + & + & - & - & - & + & + & + & 0\\
        - & - & - & - & + & + & + & 0 & + & + & +
      \end{pmatrix}.
    \end{displaymath}
    $F_7$ has $4$ variables and $11$ clauses, thus $\delta(F_7) = 11 - 4 = 7$; the hitting property is checked by visual inspection, and $F_7$ is unsatisfiable due to $8 \cdot 2^{-4} + 2 \cdot 2^{-3} + 2^{-2} = \frac 12 + \frac 14 + \frac 14 = 1$, while finally every row contains exactly one $0$, and thus $F_7$ is variable-regular of degree $10 = \nonmer(7)$.

Finally consider $A_4$ with $\delta(A_4) = 16 - 4 = 12$ and $\minvdeg(A_4) = 16$: perform four strict full subsumption resolutions on variables $1,2,3,4$, and obtain elements of $\Uclash$ of deficiency $11,10,9,8$ with min-var-degree $14,13,12,11$. \Qed
\end{prf}

\section{Conclusion and Outlook}
\label{sec:conclusion}

In this \Schrift{} we have improved the understanding of the four fundamental quantities, by supplying the lower bound $S_2 \le \maxsmarh$. The recursion defining $S_2'$ sheds also light on $S_2 = S_2'$, and we gained a deeper understanding of $S_2 = 2 a_2$. Moreover we believe (based on further numerical results)
\begin{conj}\label{con:S2exact}
  $\fa\, k \in \NN : S_2(k) = \maxsmarh(k)$.
\end{conj}
This would indeed give an unexpected precise connection of combinatorial SAT theory and elementary number theory. On the upper bound side, by Conjectures 12.1, 12.6 in \cite{KullmannZhao2010Extremal} (see Figure 1 there for a summary of the relations between the four fundamental quantities) we get:
\begin{conj}\label{con:relFCMVDH}
  $\fa\, k \in \NN : \nonmer(k) -1 \le \maxsmar(k) \le \minnonmer(k) = \minnonmerh(k)$.
\end{conj}
Recall that $\minnonmer(k) \le \nonmer(k)$; so we believe that three of the four fundamental quantities are very close to $\nonmer(k)$. This is in contrast to $\nonmer(k) - S_2(k)$ being unbounded, and indeed $S_2(k) = k+1$ for infinitely many $k$ (Corollary \ref{cor:lowbdatt}), while by Lemma \ref{lem:specdef} we also know $S_2(k) = \nonmer(k)$ for infinitely many $k$, and thus $S_2$ oscillates between the linear function $k+1$ and the quasi-linear function $\nonmer(k)$. To eventually determine the four fundamental quantities (which, if our conjectures are true, boil down to $\minnonmer$ and $\maxsmar$, while $\minnonmerh = \minnonmer$ and $\maxsmarh = S_2$), detailed investigations like those in Section \ref{sec:initvalues} need to be continued.

As $\maxsmarh(k)$ and $S_2(k)$ are closely related via (boolean) hitting clause-sets, via generalised (\emph{non-boolean}) hitting-clause-sets (see \cite{Kullmann2007ClausalFormZI,Kullmann2007ClausalFormZII} for the basic theory) we can establish a close connection to the $S_p(k)$ for all prime numbers $p$ in forthcoming work. Here $S_p(k)$ is the smallest $n \in \NNZ$ such that $p^k$ divides $n!$, as introduced in \cite[Unsolved Problem 49]{Smarandache1993OPNS}. This generalisation to (finite) domain sizes (boolean = 2) is also essential to realise the full power of the methods of this work, and to obtain applications to the field of \href{https://en.wikipedia.org/wiki/Covering_system}{covering systems of the integers}, where the relation to Boolean algebra was noticed in \cite{BergerFelzenbaumFraenkel1990Covers} (see \cite{Zeilberger2001CoveringSystems} for an introduction).

\bibliographystyle{plainurl}

\newcommand{\noopsort}[1]{}

\end{document}